\documentclass{edm_template}
\usepackage{balance}  
\usepackage{graphicx} 
\usepackage{times}    
\usepackage{url}      
\usepackage{multirow}
\usepackage{array}

\makeatletter
\def\url@leostyle{%
  \@ifundefined{selectfont}{\def\UrlFont{\sf}}{\def\UrlFont{\small\bf\ttfamily}}}
\makeatother
\def\pprw{8.5in}
\def\pprh{11in}

\usepackage[pdftex]{hyperref}

\hypersetup{
pdftitle={Learning Instructor Intervention from MOOC Forums: Early Results and Issues},
pdfauthor={LaTeX},
pdfkeywords={SIGCHI, proceedings, archival format},
bookmarksnumbered,
pdfstartview={FitH},
colorlinks,
citecolor=black,
filecolor=black,
linkcolor=black,
urlcolor=black,
breaklinks=true,
}

\begin{document}

\title{Learning Instructor Intervention from MOOC Forums: Early Results and Issues}
\numberofauthors{4}
\author{
  \alignauthor Muthu Kumar Chandrasekaran$^1$
  \vspace{.3cm}
  \alignauthor Min-Yen Kan$^1$ \and
  \vspace{.3cm}
  \alignauthor Bernard C.Y. Tan$^2$ 
  \alignauthor Kiruthika Ragupathi$^3$\thanks{
  	  This research is supported by the Singapore
	  National Research Foundation under its International Research Centre
	  @ Singapore Funding Initiative and administered by the IDM Programme
	  Office.}
  \and
    \affaddr{$^1$ Web IR / NLP Group (WING)}\\
    \affaddr{$^2$ Department of Information Systems}\\
    \affaddr{$^3$ Centre for Development of Teaching and Learning}\\
    \affaddr{National University of Singapore}\\
    \email{\{muthu.chandra, kanmy\}@comp.nus.edu.sg, \{pvotcy, kiruthika\}@nus.edu.sg}\\
}
\maketitle

\begin{abstract}
With large student enrollment, MOOC instructors face the unique
challenge in deciding when to intervene in forum discussions with
their limited bandwidth.  We study this problem of {\it instructor
  intervention}. Using a large sample of forum data culled from 61
courses, we design a binary classifier to predict whether an
instructor should intervene in a discussion thread or not. By
incorporating novel information about a forum's type into the 
classification process, we improve significantly 
over the previous state-of-the-art.

We show how difficult this decision problem is in the real world by
validating against indicative human judgment, and empirically show the
problem's sensitivity to instructors' intervention preferences. We 
conclude this paper with our take on the future research issues in 
intervention.
\end{abstract}


\category{H.3.3.}{Information Search and Retrieval}{Information filtering}
\category{K.3.1.}{Computers and Education}{Computer Uses in Education}

\section{Introduction} \label{intro}
MOOCs scale up their class size by eliminating synchronous teaching and 
the need for students and instructors to be co-located. Yet, the very 
characteristics that enable scalability of massive open online courses 
(MOOCs) also bring significant challenge to its teaching, development 
and management \cite{ferguson2014}. In particular, scaling makes it 
difficult for instructors to interact with the many
students --- the lack of interaction and feelings of isolation have
been attributed as reasons for why enrolled students drop from MOOCs
\cite{kizilcec15:_attrit_and_achiev_gaps_in_onlin_learn}.

MOOC discussion forums are the most prominent, visible artifact that
students use to achieve this interactivity.  Due to scale of
contributions, these forums teem with requests, clarifications and
social chatting that can be overwhelming to both instructors and
students alike. In particular, we focus on how to best utilize
instructor bandwidth: with a limited amount of time, which threads in
a course's discussion forum merit instructor intervention? When
utilized effectively, such intervention can clarify lecture and
assignment content for a maximal number of students, promoting the
enhancing the learning outcomes for course students.

To this end, we build upon previous work and train a binary classifier
to predict whether a forum discussion thread merits instructor
intervention or not.  A key contribution of our work is to demonstrate
that prior knowledge about forum type enhances this prediction task.
Knowledge of the enclosing forum type (i.e., discussion on {\it
  lecture}, {\it examination}, {\it homework}, etc.) improves
performance by 2.43\%; and when coupled with other known features
disclosed in prior work, results in an overall, statistically
significant 9.21\% prediction improvement.  Additionally, we show that
it is difficult for humans to predict the actual interventions (the
gold standard) through an indicative manual annotation study.

We believe that optimizing instructor intervention is an important 
issue to tackle in scaling up MOOCs. A second contribution of our 
work is to describe several issues pertinent for furthering research 
on this topic that emerge from a detailed analysis of our results. In 
particular, we describe how our work at scale details how personalized 
and individualized instructor intervention is --- and how a framework 
for research on this topic may address this complicating factor 
through the consideration of normalization, instructor roles, and 
temporal analysis.

\section{Related Work}

While the question of necessity of instructor's intervention in online
learning and MOOCs is being investigated
\cite{mazzolini2003,tomkin2014}, technologies to enable timely and
appropriate intervention are also required. The pedagogy community has
recognized the importance of instructor intervention in online
learning prior to the MOOC era (e.g., \cite{lin2009}).  Taking into
consideration the pedagogical rationale for effective intervention,
they also proposed strategic instructor postings: to guide
discussions, to wrap-up the discussion by responding to unanswered
questions, with ``Socrates-style'' follow-up questions to stimulate
further discussions, or with a mixture of questions and answers
\cite{mazzolini2007}.  However, these strategies must be revisited
when being applied to the scale of typical MOOC class sizes.


Among works on forum information retrieval, we focus on those that
focus on forum moderation as their purpose is similar to the
instructor's role in a course forum. While early work focused on
automated spam filtering, recent works shifted focus towards
curating large volumes of posts on social media platforms
\cite{backstrom2013} to distil the relevant few. Specifically, those
that strive to identify thread solvedness \cite{wang2012,kim2010} and
completeness~\cite{artzi2012} are similar to our problem.

Yet all these work on general forums (e.g., troubleshooting, or
threaded social media posts) are different from MOOC forums.  This is
due to important differences in the objectives of MOOC forums.  A
typical thread on a troubleshooting forum such as Stack Overflow is
centers on questions and answers to a particular problem reported by a
user; likewise, a social media thread disseminates information mainly
to attract attention.  In contrast, MOOC forums are primarily oriented
towards learning, and also aim to foster learning communities among
students who may or may not be connected offline.

Further, strategies for thread recommendation for students such as
\cite{yang14a} may not apply in recommending for instructors.  This
difference is partially due to scale: while the number of students and
threads are large, there are few instructors per course. In this case,
reliance on collaborative filtering using a user--item matrix is not
effective.
Learning from previous human moderation decisions~\cite{arnt2003},
therefore, becomes the most feasible approach. Prior work on MOOC
forums propose categorisation of posts \cite{ramesh14b, chaturvedi14,
  stump2013} to help instructors identify threads to
intervene. Chaturvedi {\it et al.} \cite{chaturvedi14}, the closest
related work to ours, show each of the four states of their sequence
models to predict instructor intervention to be distributed over four
post categories they infer. In this paper, we use their results for
comparison.

Different from previous works, we propose thread--level categories
rather than post--level categories, since an instructor needs to first
decide on a thread of interest. Then they need to read its content, at
least in part, before deciding whether to intervene or not. 
We make the key observation that show thread--level categories
identified as by the forum type, help to predict intervention.  

Previous work has evaluated only with a limited number of MOOC
instances.  One important open question is whether those reported
results represent the diverse population of MOOCs being taught.  In
this paper, we address this by testing on a large and diverse
cross-section of Coursera MOOC instances.



\section{Methods} \label{s:methods}

We seek to train a binary classifier to predict whether a MOOC forum
thread requires instructor intervention.  Given a dataset where
instructor participation is labeled, we wish to learn a model of
thread intervention based on qualities (i.e., features) drawn from the
dataset.  We describe our dataset, the features distilled used for our
classifier, how we obtain class labels, and our procedure for instance
weighting in the following.

\begin{figure}[t]
  \centering
  \includegraphics[width=\linewidth]{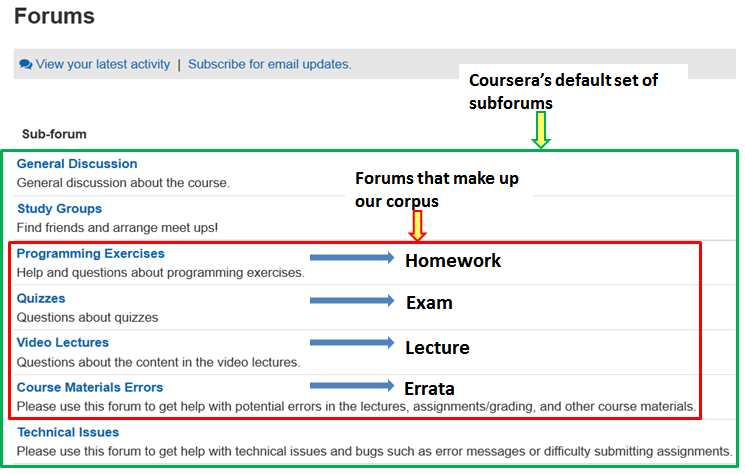}
  \caption{Typical top-level forum structure of a Coursera MOOC, with
    several forums.  The number of forums and their labels can vary
    per course.}
  \label{fig:forum}
\end{figure}

\subsection{Dataset}\label{ss:dataset}

\begin{table}[t]
  \centering
  \begin{tabular}{|l|r|r|r|r|}
    \hline
	\multirow{2}{*}{{\centering Forum type}} &
	\multicolumn{2}{p{0.3\columnwidth}|}{\centering All} &
	\multicolumn{2}{p{0.3\columnwidth}|}{\centering Intervened} \\
    \cline{2-5}
    & \# threads & \# posts & \# threads & \# posts \\
    \hline
    \hline
	\multicolumn{5}{|c|}{\textbf{D61 Corpus}} \\
	\hline
    Homework& 14,875& 127,827& 3993 & 18,637 \\
    \hline
    Lecture& 9,135& 64,906& 2,688& 10,051 \\
    \hline
    Errata& 1,811& 6,817& 654& 1,370 \\
    \hline
    Exam& 822& 6,285& 405& 1,721 \\
    \hline
    \textbf{Total} & \textbf{26,643} & \textbf{205,835} & \textbf{7,740} & \textbf{31,779} \\
    \hline 
	\hline 
	\multicolumn{5}{|c|}{\textbf{D14 Corpus}} \\
    \hline
    Homework& 3,868& 31,255& 1,385& 6,120\\
    \hline
    Lecture& 2,392& 13,185& 1,008& 3,514\\
    \hline
    Errata& 326& 1,045& 134& 206\\
    \hline
    Exam& 822& 6,285& 405& 1,721\\
    \hline
    \textbf{Total} & \textbf{7,408} & \textbf{51,770} & \textbf{2,932} & \textbf{11,561} \\
    \hline 
  \end{tabular}
  \caption{Thread statistics from our 61 MOOC Coursera dataset and the
    subset of 14 MOOCs, used in the majority of our experiments.}
  \label{tab:table1}
  \vspace{-3mm}
\end{table}

For our work, we collected a large-scale, multi-purpose dataset of
discussion forums from MOOCs. An important desideratum was to collect
a wide variety of different types of courses, spanning the full
breadth of disciplines: sciences, humanities and engineering.  We
collected the forum threads \footnote{We collected all threads and
  their component posts from four subforum categories as in
  Section~\ref{ss:dataset}. We did so, as we hypothesize that they
  would necessitate different levels of instructor intervention and
  that such interventions may be signaled by different features.} from
61 completed courses from the Coursera platform\footnote{The full list
  of courses is omitted here due to lack of space.}, from April to
August 2014, amounting to roughly 8\% of the full complement of
courses that Coursera offers\footnote{As of December 2014, Coursera, a
  commercial MOOC platform: \url{https://www.coursera.org}, hosted 761
  courses in English spanning 25 different subject areas.}.

\begin{figure}[ht]
  \centering
  \includegraphics[width=0.9\linewidth]{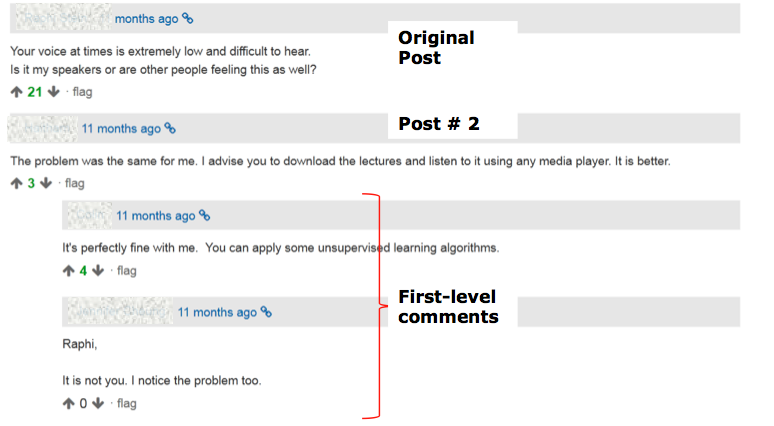}
  \caption{Coursera's forums allow threads with posts and a single level of comments.}
  \label{fig:figure2}
\end{figure}

For each course, we first assigned each forum\footnote{``Subforum'' in
  Coursera terminology.} to one of several types based on the forum's
title.  For this study we focus on threads that originated from four
prevalent types: (i) errata or course material errors, (ii) video
lectures, (iii) homework, assignments or problem sets, and (iv) exams
or quizzes (see Figure \ref{fig:forum})\footnote{Some courses had
  forums for projects, labs, peer assessment, discussion
  assignments. We omit from the collection these and other
  miscellaneous forums, such as those for general discussion, study
  groups and technical issues.}.  All 61 courses had forums for
reporting errata and discussing homework and lectures.  For more
focused experimentation, we selected the 14 largest courses within the
61 that exhibited all four forum types (denoted ``D14'' hereafter,
distinguished from the full ``D61'' dataset). Table~\ref{tab:table1}
provides demographics of both D61 and D14 datasets.  In our corpus,
there were a total of 205,835 posts including posts and comments to
posts. The Coursera platform only allows for a single level of
commenting on posts (Figure~\ref{fig:figure2}). We note that this
limits the structural information available from the forum discourse
without content or lexical analysis. We observed that posts and
comments have similar topics and length, perhaps the reason why
previous work~\cite{iri14} ignored this distinction. We have retained
the distinction as it helps to distinguish threads that warrant
intervention.

\begin{figure}[ht]
  \centering
  \includegraphics[width=0.9\linewidth]{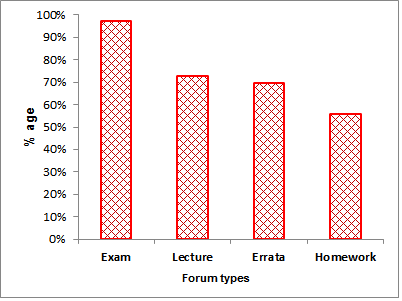}
  \caption{Thread distribution over errata, homework, lecture and exam
    forums in D14 by their \textit{intervention ratio.} }
  \label{fig:figure3}
\end{figure}

\subsection{System Design} \label{ss:sd}

From the dataset, we extract the text from the posts and comments,
preserving the author information, thread structure and posting
timestamps, allowing us to recreate the state of the forum at any
timestamp.  This is important, as we first preprocessed the dataset to
remove inadmissible information.  For example, since we collected the
dataset after all courses were completed, instructors' posts as well
as any subsequent posts in a thread need to be removed.  We also do
not use the number of votes or views in a thread as these are summary
counts that are influenced by intervention\footnote{Previous work such
  as \cite{chaturvedi14} utilize this as they have access to
  time-stamped versions of these statistics, since they use
  privately-held data supplied by Coursera for MOOCs held at their
  university.}.

We used regular expressions to further filter and canonicalize certain
language features in the forum content. We replaced all mathematical
equations by $<$MATH$>$, URLs by $<$URLREF$>$ and references to time
points in lecture videos by $<$TIMEREF$>$.  We removed stopwords,
case-folded all tokens to lowercase, and then indexed the remaining
terms and computed the product of term frequency and inverse thread
frequencies ($tf\times itf$) for term importance.  The weighted terms
form a term vector that we further normalized using the L2--norm.
Other real-valued features were max--min normalized.  Categorical
features such as the forum type were encoded as bit vectors.

Each thread is represented as bag of features consisting of terms and
specific thread metadata as disclosed below. We indicate each new
feature that our study introduces with an asterisk.
\vspace{-2mm}
\begin{enumerate}
\item [1.] Terms (unigrams);
\item [2*.] Forum type to which the thread belongs:
  Figure~\ref{fig:figure3} shows a clear difference in
  \textit{intervention ratio}, the ratio of number of threads
  intervened to those that weren't, across different forum
  types. Forum type thus emerges as a feature to use to discriminate
  threads worthy of intervention. The forum type encapsulating the
  thread could be one of homework, lecture, exam or errata.
\item [3*.] Number of references to course materials and other sources
  external to the course: includes explicit references by students to
  course materials within and outside the course e.g., \textit{slide
    4}, \textit{from wikipedia}, \textit{lecture video 7}.
\item [4*.] Affirmations by fellow students; Count of agreements made by
  fellow students in response to a post. Mostly, first posts in a
  thread receive affirmations.
\item [5.] Thread properties (Number of posts, comments, and both posts /
  comments, Average number of comments per post): expresses a thread's
  length and structural properties in terms of number of posts and
  comments posted.
\item [6.] Number of sentences in the thread: This feature intends to
  capture long focussed discussions that may be intervened more often
  than the rest.
\item [7*.] Number of non-lexical references to course materials: (number
  of URLs, references to time points in lecture videos). This feature
  is similar to course material references but includes only
  non-lexical references (Item \#1) such as URLs and time points in
  lecture videos.
\end{enumerate}
\vspace{-2mm}

Importantly, as part of the author information, Coursera also marks
instructor-intervened posts / comments.  This supplies us with
automatically labeled gold standard data for both training and
evaluating our supervised classifier.  We use threads with instructor
posts / comments as positive instances (intervened threads).  However,
we note that the class imbalance is significant: as the
instructor-student ratio is very low, typical MOOC forums have fewer
positives (interventions) than negative ones.  To counter skewness, we
weigh majority class (generally positive) instances higher than minority 
class (generally negative) instances. These weights are important parameters 
of the model, and are learned by optimizing for maximum $F_1$ over the 
training / validation set.

\section{Evaluation} \label{s:eval}

We performed detailed experimentation over the smaller D14 dataset to
validate performance, before scaling to the D61 dataset.  We describe
these set of experiments in turn.  As our task is binary
classification, we adopted L1-regularized logistic regression as our
supervised classifier in all of our experimentation.

We first investigated each of the 14 courses in D14 as 14 separate
experiments.  We randomly used 80\% of the course's threads for
training and validation (to determine the class weight parameter, $W$), and
use the remaining 20\% for testing.  Our experimental design for this
first part closely follows the previous work~\cite{chaturvedi14} for
direct comparison with their work.  We summarise these results in
Table~\ref{tab:d14}, in the columns marked ``(II) Individual'',
averaging performance over ten-fold cross validation for each course.



\begin{table*}
  \centering
  \begin{tabular}{|l||r|r||r|r|r|r||r|r|r|r|}
    \hline
    & 
    \multicolumn{2}{c||}{(I) Demographics} & 
    \multicolumn{4}{c||}{(II) Individual} & 
    \multicolumn{4}{c|}{(III) LOO-course C.V.} 
    \\
    Course &
    1. \# of Threads & 2. I. Ratio &
    1. Prec. & 2. Rec.  & 3. $F_1$ & 4. $W$ & 
    1. Prec. & 2. Rec.  & 3. $F_1$ & 4. $W$ \\
    \hline
    \hline
    ml-005& 2058 & 0.45 & 51.08&  89.19& 64.96& 2.06& 48.10& 68.63& 56.56& 2.46\\
    \hline
    rprog-003& 1123 & 0.32 & 50.77& 48.53& 49.62& 2.41& 35.88& 75.77& 48.70& 2.45\\
    \hline
    calc1-003& 965 & 0.60 & 60.98&  44.25& 51.29& 0.65& 65.42& 72.79& 68.91& 2.45\\
    \hline
    smac-001& 632 & 0.17 & 21.05& 30.77& 25.00& 5.29& 22.02& 67.93& 33.26& 2.00\\
    \hline
    compilers-004& 624 & 0.02 & 8.33&  50.00& 14.28& 37.23& 2.53& 80.00& 4.91& 2.33\\
    \hline
    maththink-004& 512 & 0.49 & 46.59& 100.00& 63.56& 2.13& 50.24& 85.48& 63.29& 2.57\\
    \hline
    medicalneuro-002& 323 & 0.76 & 100.00&  60.47& 75.36& 0.32& 75.86& 89.07& 81.94& 2.34\\
    \hline
    bioelectricity-002& 266 & 0.76 & 75.00&  54.55& 63.16& 0.34& 75.36& 82.98& 78.99& 2.41\\
    \hline
    bioinfomethods1-001& 235 & 0.55 &  56.00&  60.87& 58.33& 0.78& 59.67& 83.72& 69.68& 2.36\\
    \hline
    musicproduction-006& 232 & 0.01 & 0.00& 0.00& 0.00& 185.00& 0.52& 50.00& 1.03& 2.55 \\
    \hline
    comparch-002& 132 & 0.46 &  47.62&  100.00& 64.57& 1.56& 48.57& 83.61& 61.45& 2.37\\
    \hline
    casebasedbiostat-002& 126 & 0.20 &  13.33&  100.00& 23.53& 3.54& 24.47& 92.00& 38.66& 2.11\\
    \hline
    gametheory2-001& 125 & 0.19 & 28.57&  28.57& 28.57& 5.18& 18.27& 86.36& 30.16& 2.61\\
    \hline
    biostats-005& 55 & 0.00 &  0.00&  0.00& 0.00& 1.00& 0.00& 0.00& 0.00& 2.01\\
    \hline
    \hline
    Average & 529 & 0.36 & 39.95 & 54.80 & 41.59 & 17.68 & 37.64 & 72.74 & 45.54& 2.36\\
    Weighted Macro Avg& NA& 0.40& 45.44& 61.84& 49.04& 10.96& 42.37& 74.11& 50.56& 2.37\\
    \hline
  \end{tabular}
  \caption{Individual course results for each course in the D14
    dataset.  Weights $W$ weigh each +ve class instance $w$ times as
    much as a --ve class instance. Performance varies with large variations in Intervention ratio (I-ratio) and \# of threads.}
  \label{tab:d14}
\end{table*}

The results show a wide range in prediction performance. This casts
doubt on the portability of the previously published work \cite{chaturvedi14}. 
They report a baseline performance of $F_1\approx25$ on both their courses each 
having an intervention ratio $\approx0.13$\footnote{Based on test data figures \cite{chaturvedi14} 
had disclosed in their work}. In contrast, our results show the instability of the 
prediction task, even when using individualized trained models. Nevertheless, 
on average our set of features performs better on $F_1$ by at least 10.15\%.\footnote{Due 
to non-availability of experimental data, we can only claim a 10.15\% improvement over the highest $F_1$ they reported, 35.29.}

\begin{figure}[t]
  \centering
  \includegraphics[width=0.9\linewidth]{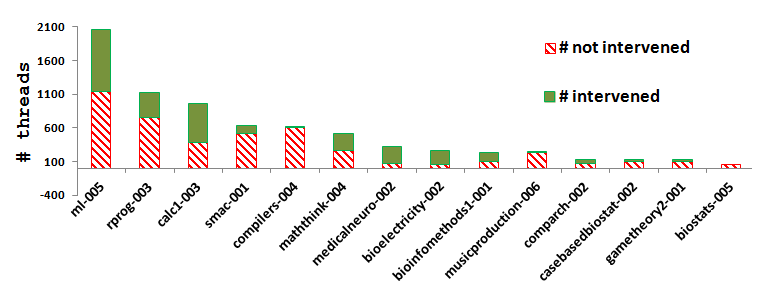}
  \caption{Thread distribution over the errata, homework, lecture and
    exam forums in D14.  Corresponds to numeric data in
    Table~\ref{tab:d14}.}
  \label{fig:figure1b}
\end{figure}

We observe the true intervention ratio correlates to performance, when
comparing Columns I.2 and II.3 ($\rho=0.93$). We also see that intervention ratio
varies widely in our D14 dataset (Figure~\ref{fig:figure1b}). This 
happens to also hold for the larger D61 dataset. In some courses,
instructors intervene often (76\% for medicalneuro-002) and in some
other courses, there is no intervention at all (0\% for biostats-005).

To see whether the variability can be mitigated by including more
data, we next perform a leave-one-course-out cross validation over the
14 courses, shown in ``Columns (III) LOO-course C.V.''.  {\it I.e.},
we train a model using 13 courses' data and apply the trained model to
the remaining unseen test course. While not strictly comparable with (II), 
we feel this setting is more appropriate, as it: allows training to scale;  
is closer to the real scenario discussed in Section \ref{s:framework}, Item 4.


Separately, we studied the effectiveness of our proposed set of
features over the D14 dataset.  Table~\ref{tab:features} reports
performance averaged over all 14 courses weighted by its proportion in
the corpus.  In the top half of the table, we build Systems 1--7 by
cumulatively adding in features from the proposed list from
Section~\ref{ss:sd}.  Although the overall result in Row~7 performs
$\sim5\%$ better than the unigram baseline, we see that the classifier
worsens when the count of course references are used as a feature
(Row~2). Other rows all show an additive improvement in $F_1$,
especially the forum type and non-lexical reference features, which
boost recall significantly.


The performance drop when adding in the number of course references
prompts us to investigate whether removing some features from the full
set would increase prediction quality.  In the bottom half of
Table~\ref{tab:features}, we ablate a single feature from the full
set.



Results show that removing forum type, number of course references and thread 
length in a thread all can improve performance. Since the different rows of 
Table~\ref{tab:features} are tested with weights $W$ learnt from its own training 
set the changes in performance observed are due to the features and the learnt 
weight. When we tested the same sequence with an arbitrary constant weight we 
observed all features but Course\_Ref improved performance although not every 
improvement was significant.

\begin{table}
  \centering
  \begin{tabular}{|l|r|r|r|}
    \hline
    Feature &
    Precision &
    Recall &
    $F_1$ \\ 
    \hline
    \hline
    1. Unigrams & 41.98&  61.39&  45.58 \\ 
    \hline
    2. (1) + Forum Type & 41.36& 69.13& 48.01 \\ 
    \hline
    3. (2) + Course\_Ref & 41.09& 66.57& 47.22\\ 
    \hline
    4. (3) + Affirmation & 41.20& 68.94& 47.68\\ 
    \hline
    5. (4) + T Properties & 42.99& 70.54& 48.86\\ 
    \hline
    6. (5) + Num Sents & 43.08& 69.88& 49.77\\ 
    \hline   
    7. (6) + Non-Lex Ref & 42.37& 74.11& 50.56\\ 
    \hline 
    \hline
    8. (7) -- Forum Type & 41.33& 83.35& 51.16 \\ 
    \hline
    9. (7) -- Course Ref & 45.96& 79.12& 54.79 \\ 
    \hline
    10. (7) -- Affirmation & 42.59& 71.76& 50.34 \\ 
    \hline
    11. (7) -- T Properties & 40.62& 84.80& 51.35 \\ 
    \hline
    12. (7) -- Num Sents & 42.37& 73.05& 49.32 \\ 
    \hline
    13. (7) -- Non-Lex Ref & 43.08& 69.88& 49.77\\ 
    \hline 
    
  \end{tabular}
  \caption{Feature study.  The top half shows performance as
    additional features are added to the classifier.  Ablation tests
    where a single feature is removed from the full set (Row~7) are
    shown on the bottom half.  Performance given as weighted
    macro-average over 14 courses from a leave-one-out cross course
    validation over D14. 
  }
  \label{tab:features}
\end{table}
\vspace{-1mm}


Using the best performing feature set as determined on the D14
experiments, we scaled our work to the larger D61 dataset.
Since a leave-one-out validation of all 61 courses is time consuming 
we only test on the each of the 14 courses in D14 dataset while training 
on the remaining 60 courses from D61. We report a \textbf{weighted averaged 
$F_1$ = 50.96 (P = 42.80; R = 76.29)} which is less than row 9 of Table~\ref{tab:features}. 
We infer that scaling the dataset by itself doesn't improve performance since 
$W$ learnt from the larger training data no longer counters the class imbalance 
leaving the testset with a much different class distribution than the training set.



\subsection{Upper bound}
\label{ss:upperbound}
The prediction results show that forum type and some of our
newly-proposed features lead to significant improvements. However, we
suspect the intervention decision is not entirely objective; the
choice to intervene may be subjective.  In particular, our work is
based on the premise that correct intervention follows the historical
pattern of intervention (where instructors already intervene), and may
not be where general pedadogy would recommend prediction.  We
recognize this as a limitation of our work.

To attempt to quantify this problem, we assess whether peer
instructors with general teaching background could replicate the
original intervention patterns.  Three human instructors\footnote{The
  last three authors, not involved in the experimentation: two
  professors and a senior pedagogy researcher.}  annotated 13 threads
from the musicproduction-006 course.  We chose this course to avoid
bias due to background knowledge, as none of the annotators had any
experience in music production. This course also had near zero
interventions; none of the 13 threads in the sample were originally
intervened by the instruction staff of the course.

They annotated 6 exam threads and 7 lecture threads. We found that 
among exam threads annotators agreed on 5 out of 6 cases. Among lecture 
threads at least two of three annotators always agreed.  On 4 out of 7 
cases, all three agreed. The apparently high agreement could be because 
all annotators chose to intervene only on a few threads. This corresponds 
to a averaged interannotator agreement of $k=0.53$. 
The annotators remarked that it was difficult to make judgements, that
intervention in certain cases may be arbitrary, especially when expert
knowledge would be needed to judge whether factual statements made by
students is incorrect (thus requiring instructor intervention to
clarify). As a consequence, agreement on exam threads that had questions 
on exam logistics had more agreement at $k=0.73$.

While only indicative, this reveals the subjectiveness of
intervention. Replicating the ground truth intervention history may
not be feasible -- satisfactory performance for the task may come
closer to the interannotator agreement levels: i.e., $k=0.53$ corresponding
to an $F_1$ of 53\%.  We believe this further validates the
significance of the prediction improvement, as the upper bound for
deciding intervention is unlikely to be 100\%.


\newpage
\section{Discussion} \label{sec:disc}

From handling the threads and observing discussion forum interactions
across courses, several issues arise that merit a more detailed
discussion.  We discuss each in turn, identifying possible actions
that may mollify or address these concerns.  Specifically:


\vspace{-4mm}

\begin{enumerate}
\item The number of threads per course varies significantly.

\item Intervention decisions may be subjective.

\item Simple baselines outperform learned systems.

\item Previous experimental results are not replicable.
\end{enumerate}
\vspace{-4mm}

{\bf Issue 1: Variation in the number of threads.}  We observed
significant variation in the number of threads in different courses,
ranging from tens to thousands. Figure~\ref{fig:figure1b} shows thread distribution
over the D14 dataset for the errata, homework, lectures and exam
forums; a similar distribution held for the larger D61 dataset. These
distributions are similar to those reported earlier in the large
cross-course study of~\cite{iri14}.
The difference in number of threads across courses is due to a multitude
of factors. These include number of students participating, course
structure, assignment of additional credits to participating students,
course difficulty, errors in course logistics and materials, etc.

When performing research that cuts across individual MOOCs, this issue
becomes important.  As we saw, using simple averaging on a per-course
basis equates to a macro-averaging: putting each course on par in
importance. However, when the decision unit is at the thread (as in
our task), it makes more sense to treat individual threads at parity.
In such cases, normalization at the thread level (analogous to
micro-averaging) may be considered. Such thread-level normalization
can affect how we weight information from each course when training in
aggregate over data from multiple courses: courses with many threads
should carry more weight in both training and evaluation.

\vspace{-2mm}

{\bf Issue 2: Intervention decisions may be subjective.} Instructor
policy with respect to intervention can markedly differ. Instructors
may only intervene in urgent cases, preferring students to do peer
learning and discovery. Others may want to intervene often, to
interact with the students more and to offer a higher level of
guidance. Which policy instructors adopt varies, as best practices for
both standard classroom and MOOC teaching have shown both advantages
and disadvantages for \cite{mazzolini2003, mackness2010}.

Instructors can also manifest in different roles.  In Coursera, posts
and comments marked as instructor intervened can come from actual
instructor intervention as well as participation by helpers, such as
community teaching assistants (CTAs).  
We observe courses with CTAs where CTAs have a higher intervention
rate. We hypothesize that such factors decreases agreements.

This plays out in our datasets.  We observe that intervention is not
always proportional to the number of threads in the course. Some
courses such as compilers-004~(see Figure \ref{fig:figure1b}) has relatively fewer
number of threads than other large courses.  Yet its intervention rate
is noticeably low. This suggests that other factors inform the
intervention decision.  Handling this phenomenon in cross-course
studies requires an additional form of normalization.

To normalize for these different policies we can upweight (by
oversampling) threads that were intervened in courses with fewer
interventions.  We can continue to randomly oversample a course's
intervened threads until its {\it intervention density} reaches the
dataset average.  Note this normalization assumes that
the few threads intervened in course with relatively low intervention
density are more important; that the threads intervened for a similar
high intervention density course would be a proper superset.

Even when a policy is set, intervention decisions may be subjective
and non-replicable.  Even with our cursory annotation of a course to
determine an upper bound for intervention shows the potentially large
variation in specific intervention decisions.  We believe that
automated systems can only approach human performance when such
decisions can be subjective.  As such, the upper bound for performance
(cf Section~\ref{ss:upperbound}) should not be construed as the single gold
standard; rather, prediction performance should be calibrated to human
performance levels.





\begin{figure*}
   \centering
   \includegraphics[width=\linewidth]{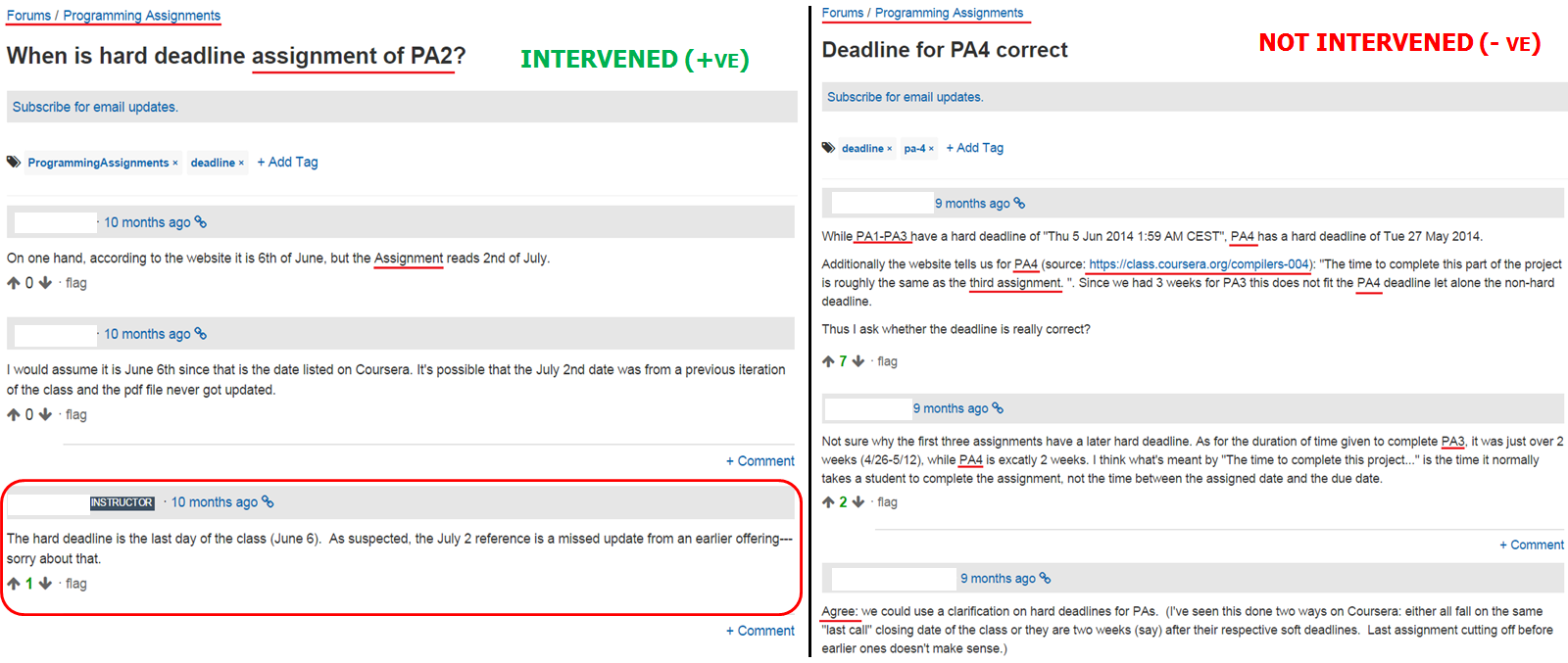}
   \caption{Interventions are, at times, arbitrary. We show two
     threads from {\tt compilers-001} with similar topics, context,
     and features that we model (red underline). Yet only one of
     them is intervened (circled in red).}
   \label{fig:arbitrary}
 \end{figure*}

{\bf Issue 3: Simple baselines outperform learned models.}  We also
compared our work with a simple baseline that predicts all threads as
needing instructor intervention.  This baseline does no work --
achieving 100\% recall and minimal precision -- but is very
competitive, outperforming our learned models for courses with high
levels of intervention (see Table~\ref{tab:100R}).
Diving deeper into the cause, we attribute this difference to the
subjective nature of interventions and other extraneous reasons
(bandwidth concerns) resulting in high false positive rates.
That is, given two threads with similar set of features, one may be
intervened while the other is not (e.g., Figure~\ref{fig:arbitrary}).
This makes the ground truth and the evaluation less reliable. An 
alternative evaluation model might be to assign a confidence score 
to a prediction and evaluate the overlap between the high confidence 
predictions and the ground truth interventions.
\vspace{-1mm}

\begin{table}
\small
\centering
\begin{tabular}{|l|r|r||r|r|}
    \hline
     &
    \multicolumn{2}{c||}{Individual} &
    \multicolumn{2}{c|}{D14} \\
    \cline{2-5}
	Course &
	$F_1$ &
	$F_1$@100R&
	$F_1$ &
	$F_1$@100R\\
	\hline	
        \hline	
	ml-005& 64.96& 63.79& 72.35& 61.83\\
	\hline
	rprog-003& 49.62& 47.39& 48.55& 49.31\\
	\hline
	calc1-003& 51.29& 74.83& 70.63& 75.33\\
	\hline
	smac-001& 25.00& 34.67& 34.15& 29.28\\
	\hline
	compilers-004& 14.28& 3.28& 4.82& 4.75\\
	\hline
	maththink-004& 63.56& 63.08& 61.11& 65.49\\
	\hline
	medicalneuro-002& 75.36& 88.66& 78.06& 85.67\\
	\hline
	bioelectricity-002& 63.16& 86.84& 80.10& 85.84\\
	\hline
	bioinfomethods1-001& 58.33& 67.65& 69.40& 71.07\\
	\hline
	musicproduction-006& 0.00& 4.35& 1.09& 1.72\\
	\hline
	comparch-002& 64.57& 55.56& 60.49& 63.21\\
	\hline
	casebasedbiostat-002& 23.53& 14.81& 38.71& 34.25\\
	\hline
	gametheory2-001& 28.57& 45.16& 27.12& 30.56\\
	\hline
	biostats-005& 0.00& 0.00& 0.00& 0.00\\
\hline	
\hline	
	Average& 41.59& 46.43& 45.18& 47.09\\
	Weighted Macro Avg& 49.04& 51.51& 54.79& 53.22\\
\hline	
\end{tabular}
\caption{Comparison of $F_1$ in Table~\ref{tab:d14} with those of a
  na\"{\i}ve baseline that classifies every instance as +ve --
  resulting in 100\% recall.}
\label{tab:100R}
\end{table}


\vspace{-1mm}
{\bf Issue 4: Previous results are not replicable.} From earlier work
\cite{chaturvedi14}, intervention prediction seemed to be
straightforward task where improvement can be ascribed to better
feature engineering.  However, as we have discovered in our datasets,
the variability in instructor intervention in MOOCs is high, making
the application of such previously published work to other MOOCs
difficult.  This is the perennial difficulty of replicating research
findings.  Findings from studies over a small corpus with select
courses from specific subject categories may not generalise.
Published findings are not verifiable due to restricted access to
sensitive course data. The problem is acute for discussion forum data
due to privacy and copyright considerations of students who have
authored posts on those forums. 

\vspace{-1mm}
The main challenge is to provision secured researcher access to the
experimental data.  Even in cases where researchers have access to
larger datasets, such prior
research~\cite{anderson2014,chaturvedi14,ramesh13a,ramesh14a,ramesh14b,wen14}
have reported findings on each course separately (cf
Table~\ref{tab:d14} ``(II) Individual''), shying away from compiling
them into a single dataset in their study.  Bridging this gap requires
cooperation among interested parties.  The shared task model is one
possibility: indeed, recently Rose {\it et al.} \cite{rose2014}
organised a shared task organised to predict MOOC dropouts over each
week of a MOOC.  To effectively make MOOC research replicable, data
must be shared to allow others to follow and build on published
experimentation.  Similar to other communities in machine learning and
computational linguistics, the community of MOOC researchers can act
to legislate data sharing initiatives, allowing suitably anonymized
MOOC data to be shared among partner institutes.

\vspace{-1mm}
We call for the community to seize this opportunity to make research
on learning at scale more recognizable and replicable.  We have gained
the endorsement of Coursera to launch a data-sharing initiative with
other Coursera-partnered universities.  While we recognize the
difficulties of sharing data from the privacy and institutional review
board perspectives, we believe that impactful research will require
application to a large and wide variety of courses, and that
restricting access to researchers will alleviate privacy concerns.
\vspace{-1mm}
\section{A Framework for Intervention Research} \label{s:framework}

We have started on the path of instructor intervention prediction,
using the task formalism posed by previous work by Chaturvedi {\it et
  al.} \cite{chaturvedi14}: the binary prediction of whether a forum
discussion thread should be intervened, given its lexical contents and
metadata.  While we have improved on this work and have encouraging
results, this binary prediction problem we have tackled is overly
constrained and does not address the real-world need for intervention
prediction. We outline a framework for working towards the real-world
needs of instructor intervention.

We thus propose a framework for investigation that iteratively relaxes
our problem to take into account successively more realistic aspects
of the intervention problem, with the hope of having a fieldable,
scalable, real-time instructor intervention tool for use on MOOC
instructors' dashboard as an end result.

  \textbf{1. Thread Ranking.} We posit that different types of student
  posts may exhibit different priorities for instructors. A 
  recommendation for intervention should also depend on thread
  criticality. For example, threads reporting errors in the course
  material may likely be perceived as critical and hence should be
  treated as high-priority for intervention.  Even with designated
  errata forums, errata are reported in other forums, sometimes due to
  the context -- {\it e.g.}, when a student watches a video of a
  lecture, it is natural for him to report an error concerning it in
  the lecture forum, as opposed to the proper place in the errata
  forum.  Failure to address threads by priority could further increase the
  course's dropout rate, a well-known problem inherent to MOOCs
  \cite{clow13}.  Thread ranking can help to address this problem to
  prioritize the threads in order of urgency, which the na\"{\i}ve,
  always classifying all instances as positive, baseline system cannot
  perform.
\vspace{-1mm}

\textbf{2.  Re-intervention.} Threads can be long and several related
  concerns can manifest within a single thread, either by policy or by
  serendipity.  Predicting intervention at the thread level is
  insufficient to address this.  A recommendation for intervention has
  to consider not only those threads that had been newly-created but
  also if older threads that had already been intervened require
  further intervention or \textit{reintervention}. In other words,
  intervention decision needs to be made in the light of newly posted
  content to a thread.  We can change the resolution of the
  intervention prediction problem to one at the post level, to capture
  re-interventions; {\it i.e.}, when a new post within a thread
  requires further clarification or details from instructor staff.
\vspace{-1mm}

\newpage
\textbf{3. Varying Teaching Roles.} MOOCs require different instruction
  formats than the traditional course format.  One evolution of the
  MOOC teaching format to adapt to the large scale is to recruit
  community teaching assistants (CTA)s.  Community TAs are volunteer
  TAs recruited by MOOC platforms including Coursera based on their
  good performance in the previous iteration of the same MOOC.  CTAs,
  traditional Teaching Assistants and technical staff are all termed
  as ``staff'' within the Coursera system.  Currently, Coursera
  only marks threads with a ``staff replied'' marking, which we use
  directly in our training supervision in this paper. At a post level, 
  those posted by CTAs, instructor and technical staff are marked appropriately.
\vspace{-1mm}

  We hypothesize that that these various roles differ in the quantum
  of time and effort, and type of content that they provide in
  answering posts that they contribute on a forum.  It will be
  important to consider the role of the user while recommending
  threads to intervene, as the single problem of intervention may lead
  to $n$ separate triaging problems for the $n$ staff types or
  individual instructors that manage a MOOC.
\vspace{-1mm}

\textbf{4. Real-time.} In the real world, a system needs to be predicting 
  intervention in real-time; as new posts come into a course's forum.
  With ranking, we can decide when to push notifications to the
  instruction staff, as well as those less urgent that can be viewed
  at leisure on the instructor's MOOC dashboard.  
\vspace{-1mm}

  With the timestamp metadata in the dataset, we have a transaction
  log.  This allows us to easily simulate the state of a MOOC by
  ``rewinding'' the state of the MOOC at any time $t$, and make a
  prediction for a post or thread based on the current state.
\vspace{-1mm}

  This half-solves the problem. For real-world use, we also need to
  do online learning, by observing actual instructor intervention and
  adopting our system for the observed behavior.  We feel this will be
  important to learn the instructor's intervention preference, as we
  have observed the variability in intervention per course, per
  instructor.
\vspace{-1mm}

In our work, we focus only on the {\it instructor's view}, however
this set of problems also has an important dual problem set: that of
the {\it student's view}.  We believe that solving both problems will
have certain synergies but will differ in important ways.  For
example, solving the student's view will likely have a larger peer and
social component than that for instructors, as MOOCs develop more
social sensitivity.

\section{Conclusion} \label{conc}
We describe a system for predicting instructor intervention in MOOC
forums. Drawing from data over many MOOC courses from a wide variety
of coursework, we devise several novel features of forums that allow
our system to outperform the state-of-the-art work on an average 
by a significant margin of 10.15\%. In particular, 
we find that knowledge of where the thread originates from 
(the forum type -- whether it appears in a {\it lecture}, 
{\it homework}, {\it examination} forum) alone informs the 
intervention decision by a large 2\% margin.

While significant in its own right, our study also uncovers issues
that we feel must be accounted for in future research.  We have
described a framework for future research on intervention, that will
allow us to account for additional factors -- such as temporal
effects, differing instructor roles -- that will result in a ranking
of forum threads (or posts) to aid the instructor in best managing her
time in answering questions on MOOC forums.

Crucially, we find the amount of instructor intervention is widely
variable across different courses.  This variability undermines the
veracity of previous works and shows that what works on a small scale
may not hold well in large, cross-MOOC studies.  Our own results show
that for many courses, simple baselines work better than supervised
machine learned models when intervention ratios are high.  To allow
the replicability of research and to advance the field, we believe
that MOOC-fielding institutions need to form a data consortium to make
MOOC forum data available to researchers.

\section{Acknowledgments}

The authors would like to thank Snigdha Chaturvedi and her co-authors
for their help in answering detailed questions on the methodology of
their work.

\bibliographystyle{abbrv}
\bibliography{edm2015}
\vspace{-3mm}
\balance
\end{document}